\documentclass[preprint,prd,tightenlines,nofootinbib,superscriptaddress]{revtex4-1}

\usepackage{color}
\usepackage{graphicx}
\definecolor{red}{rgb}{1,0,0}
\def\lesssim{\ \hbox{\raise 2pt \hbox{$<$} \kern -13pt
                     \lower 3pt \hbox{$\sim$}}\ }
\def\greatersim{\ \hbox{\raise 2pt \hbox{$>$} \kern -13pt
                     \lower 3pt \hbox{$\sim$}}\ }

\input epsf.tex
\def\desepsf(#1 width #2){\epsfxsize=#2 \epsfbox{#1}}

\usepackage{amsmath,bm}
\begin{document}

\vspace*{1.4 cm} 

\title{The CASTOR calorimeter at the CMS experiment}
\author{P.\ Gunnellini\footnote{paolo.gunnellini@desy.de}}
\affiliation{Deutsches Elektronen Synchrotron, Notkestr.85, D-22603 Hamburg}

\begin{center}
{\small On behalf of the CMS collaboration} 
\end{center}
\begin{center}
$41^{st}$ ITEP Physics School, Moscow 2013
\end{center}
\hspace*{12.9 cm} {\small } 

\vspace{3cm}
\begin{abstract}
The CASTOR Calorimeter at the CMS experiment is an electromagnetic/hadronic calorimeter which covers the very forward region of the detector (-6.6 $<$ $\eta$ $<$ -5.2). CASTOR is a Cherenkov sampling calorimeter, consisting of quartz and tungsten plates, with an overall depth of 10 interaction lengths, able to detect penetrating cascade particles. It is segmented in 16 transversal and 14 longitudinal sections. Surrounding the beam pipe, its design is determined by space constraints and restricted to materials which tolerate a high radiation level. In this presentation we report on the operational experience and measurements with the CASTOR calorimeter during the 2010 data taking at the LHC, with proton-proton and heavy ion collisions. An overview of the broad physics program which can be accessed with CASTOR, as well as the status of published and ongoing physics analyses and detector studies are presented.
\end{abstract} 


\maketitle

\section{Introduction} 
An important part of the physics program at the Large Hadron Collider (LHC) involves the forward rapidities, due to particles produced at very low angles with respect to the beam line. The interest of the forward physics is very wide and significant: in fact, it offers a broad two-fold program \cite{forwardphysics}. On one hand, it helps to understand the background of more rare processes, for any kind of discovery channels, by giving light to the modelling of the underlying event in both \textit{pp} or \textit{PbPb} collisions; on the other hand it opens up to several discoveries related to the proton structure or the parton evolution. In particular, dijet-measurements or rapidity gap events where an activity in the detector is observed only in regions highly separated in pseudorapidity (see fig. \ref{physics}a), are the benchmarks for the study of the soft and hard diffraction and they are accessible mainly by looking at the forward region. Very interesting is also the whole phenomenology of the low-x QCD where a pair of partons with big difference in longitudinal momentum fraction interacts: it allows to give insights on the parton distribution functions, it enables to scan different regimes of the parton evolution (see fig. \ref{physics}b) providing, in case, a direct measurement of the gluon saturation and it is relevant to study the contribution of the multiple parton interactions (MPI).\\
A more detailed summary of the forward physics program can be found in \cite{forwardphysics}.\\
The Compact Muon Solenoid (CMS) experiment at the LHC at CERN is well equipped to investigate the physics at the forward region with several subdetectors based on calorimetry detection, aiming at the achievement of the goals, previously described. \\
A brief overview of the CMS experiment is given in section 2 with a wider description of the CASTOR calorimeter that, being  the most forward subdetector, has to face several challenges due to the harsh and highly radiated environment where it is located. In section 3, the first measurements performed with CASTOR and the physics implications of the results are reported.

\section{The CMS experiment and the CASTOR calorimeter} 
A very detailed description of the CMS detector can be found in \cite{CMS}. Its structure is shown in figure \ref{detectors}a with all the subdetectors that are described below.\\
The central feature of the Compact Muon Solenoid (CMS) experiment at the LHC is a superconducting solenoid of 6 metres internal diameter, that provides a magnetic field of 3.8 T for the measurement of the momentum of the charged particles. Within the field volume are the silicon pixel and strip tracker, the crystal electromagnetic calorimeter (ECAL) and the brass/scintillator hadron calorimeter (HCAL). Muons are measured in gas-ionization detectors embedded in the steel return yoke. The CMS trigger system is well described and treated in \cite{TDR1} and \cite{Trigger}. In addition to the barrel and endcap detectors, CMS has extensive forward calorimetry. The hadronic forward (HF) calorimeters cover the region 2.9 $<$ $\eta$ $<$ 5.2. They consist of iron absorbers and embedded radiation-hard quartz photomultiplier tubes. Calorimeter cells are formed by grouping bundles of fibres. Clusters of these cells (e.g. 3 $\times$ 3 grouping) form a calorimeter tower. There are 13 towers in $\eta$, each with a size given by $\Delta\eta$ $\approx$ 0.175, except for the lowest- and highest-$|\Delta\eta|$ with $\Delta\eta$ $\approx$ 0.1 and $\Delta\eta$ $\approx$ 0.3, respectively. The azimuthal segmentation of all towers is $10^{\circ}$, except for the one at highest $|\eta|$, which has $\Delta\phi$ = $20^{\circ}$.\\
Even more forward angles, -6.6 $<$ $\eta$ $<$ -5.2, are covered by the CASTOR calorimeter (pictured in fig.\ref{detectors}b), though only on the negative longitudinal side of CMS: it extends thus the CMS acceptance to the very forward region, for a total coverage of -6.6 $<$ $\eta$ $<$ 5.2. This calorimeter is made of quartz plates embedded in tungsten absorbers, providing a fast collection on the Cherenkov light. The collected light is detected using fine-mesh Hamamatsu R5505 photomultiplier tubes, which allow operation under up to 0.5 T magnetic field if the field direction is within $\pm$$45^{\circ}$ with respect to the photomultiplier axis \cite{testbeam}. The calorimeter is segmented in 16 $\phi$-sectors and 14 z-modules. The first two modules have half the depth of the others and serve to detect electromagnetic showers. The full calorimeter has a depth of 10.5 interaction lenghts. The performance of the CASTOR calorimeter was studied in a test beam environment \cite{performancestudies}.\\
The main challenge in the operation of CASTOR is the very special location at about $z=14.3 m$ from the interaction point, close to the beam pipe and surrounded by massive shields. This requires a very compact form of the detector, with one of the consequences being that the 224 photomultiplier tubes (PMTs) are mounted directly on the detector, less than 30 cm away from the LHC beam. The PMTs are thus exposed to high-radiation levels and strong fringe magnetic fields.\\
In particular, the complicated magnetic field configuration at the location of CASTOR is caused by the fact, that the massive shields that surround CASTOR meet in proximity of its center (around module 7), producing an air gap of 40 mm between them. The absolute value of the magnetic field flux measured at this region does not exceed 0.2 T, however, the direction of the field varies strongly. This results in totally suppressed responses of the PMTs located around the gap in the shielding, as demonstrated in figure \ref{magnfieldshifts}a. Therefore, with modules 6 to 8 suffering from the magnetic field and modules 9 to 14 collecting only a small fraction of hadronic showers, the ongoing analyses, besides to the ones described in section 3 are restricted to the modules 1 to 5. This corresponds to a reduction of the calorimeter depth from 10.5 to 3.2 hadronic interaction lenghts. It has been checked from simulation that about 80$\%$ of the energy deposited in CASTOR in inclusive events is contained in the first 5 modules.\\
Another consequence of the strong remnant fields in the forward region of the CMS detector is that the CASTOR calorimeter slightly shifts when the CMS solenoid is switched on. Figure \ref{magnfieldshifts}b shows the location of the two CASTOR halves as measured by position sensors that serve to monitor movements that may harm the beam pipe. The largest shift is found to be approximately 12 mm. This results in some $\phi$ sectors to move to more central rapidity, covering the range between -6.3 $<$ $\eta$ $<$ -5.13. The effect of these movements is considered in the estimate of systematic uncertainties and a strong effort is being put on the software simulation in order to implement this shift.\\
The response of individual CASTOR cells has been equalized using a sample of beam halo muon events. An absolute calibration of 0.015 GeV/fC, with an uncertainty of $\pm$30$\%$, is obtained from a Monte Carlo based extrapolation of the $\eta$ dependence of the energy density per unit of pseudorapidity measured in the HF calorimeter to the CASTOR acceptance \cite{Calibration}. Even though this result is found to be with test beam measurements, work is in progress to perform a physics-based absolute calibration: events with two balanced objects, one reconstructed in CASTOR and the other one in a well calibrated region of the CMS detector, can be exploited for this aim. Good candidates are di-jet events, Z+jets with the Z decaying leptonically or di-electron events in ultra-peripheral collisions.  

\section{Study of the underlying event in proton-proton and lead-lead collisions} 
The study of the underlying event in \textit{pp} and \textit{PbPb} collisions has been performed with the CMS experiment by measuring the energy density in the CASTOR $\eta$-region.\\
By comparing the measurements to several predictions that use different physics inputs, it is possible to extract important conclusions in order to exclude or constrain the models currently used for the underlying event.\\
The underlying event activity for \textit{pp} collision in the forward rapidity has been measured through the ratio of the energy density $dE/d\eta$ between events with a charged particle jet produced at central rapidity (-2 $<$ $\eta$ $<$ 2) and inclusive events where no requests in the central region have been asked. This quantity was measured as a function of the charged particle jet transverse momentum at three different center-of-mass energies ($\sqrt{s}$ = 0.9, 2.76 and 7 TeV) \cite{ppforward}. This measurement is particular sensitive to the behaviour of the beam remnants and the contribution of the MPIs.\\
Two different sets of Monte Carlo generators were used in order to compare the measurement with their predictions: besides to the standard ones, Pythia \cite{pythref} and Herwig \cite{herwref}, based on the description of the hard scattering at the accelerators, cosmic-ray based Monte Carlo generators, like  HYDJET \cite{HYDJET}, AMPT \cite{AMPT}, EPOS \cite{epos}, SIBYLL \cite{sibyll} and QGSJET \cite{QGSJET}, which are aimed to describe the high-energy interactions between the cosmic rays and the particles of the atmosphere, were also compared.\\
The data corrected to the stable particle level and the Monte Carlo predictions are shown in figure \ref{PPmeasurement1} and \ref {PPmeasurement2}, for respectively $\sqrt{s}$ = 0.9 and 7 TeV. They show, in particular, a nice agreement for both energies between the data and the Pythia Z2* and 4C, and the Herwig~2.5 predictions, while the old D6T Pythia tune predicts too much MPI and fails to describe the data. None of the cosmic ray models is able to give a good representation of the data.\\
Another important feature that can be extracted from this measurement is the different shape of the energy density ratios at different $\sqrt{s}$: as shown in figures \ref{PPmeasurement1} and \ref{PPmeasurement2}, at 0.9 TeV the curve saturates at values below 1 at $p_T$ $>$ 5 GeV, while at 7 TeV it goes constantly above 1. These shapes are due to two competing phenomena: on one hand, in presence of a hard scattering, less energy is available for the underlying event and the ratio tends to be below 1; on the other hand, the contribution of the MPIs to the underlying event grows with the increase of the overlap between the two interacting protons and the hardness of the scattering, making the ratio higher. At low $\sqrt{s}$, the first effect is dominant, and the relevance of the second one increases with the collision energy.\\

The energy density has been also measured with the CMS detector as a function of $\eta$ in \textit{PbPb} collisions at 2.76 TeV center-of-mass energy \cite{leadleadforward}.\\
This measurement has been performed in the whole pseudorapidity coverage, including the CASTOR region, looking at different ranges of centrality. The centrality of a collision is defined by the impact parameter of the two interacting nuclei. The results are shown in figure \ref{leadlead}.\\ 
Different Monte Carlo generator predictions were compared to the data but none of them is able to describe the data in the whole phase space or in all centrality bins. Still work needs to be done to tune the physics models that are currently used to describe heavy-ions collisions.

\section{Summary and conclusions}
In this paper, a brief overview of the topics related to the forward physics and a description of the CASTOR calorimeter at the CMS experiment is provided. Some of the most actual detector issues and challenges have been also treated. In the last section, two of the most relevant measurements performed with CASTOR and some physics considerations have been given in order to interprete the observations.

\section*{Acknowledgements}
I would like to thank the organizers of the $41^{st}$ ITEP Physics school for their kindness and professionality and for giving me the possibility to present this interesting piece of work in that stimulating environment. Special thanks also to Hannes Jung and Kerstin Borras for their nice supervision and the warm incitement for the application at the school and the CASTOR workgroup that performed the analysis I presented.

\newpage

\begin{figure}[htbp]
\begin{center}
\includegraphics[width=8.5cm]{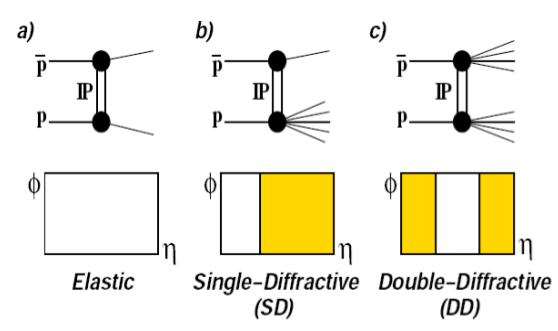}
\includegraphics[width=7.5cm]{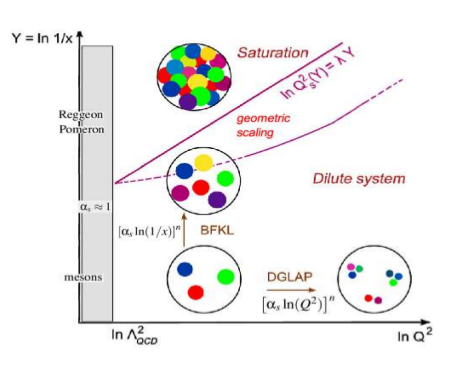}
\caption{}  
\label{physics}
\end{center}
\end{figure}

{\bf Fig.~\ref{physics}}. Left: Event topologies in a pseudorapidity vs azimuthal angle plane: the shaded and empty areas represent respectively particle production and gap regions. Right: QCD log(1/x)-$Q^2$ plane to show the different parton evolution regimes (DGLAP, BFKL, saturation.

\begin{figure}[htbp]
\begin{center}
\includegraphics[width=8cm]{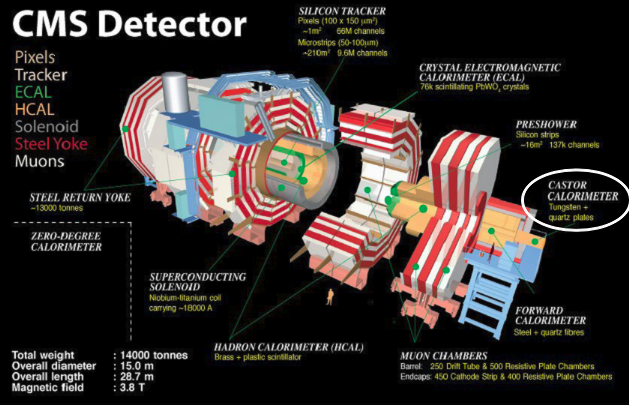}
\includegraphics[width=8cm]{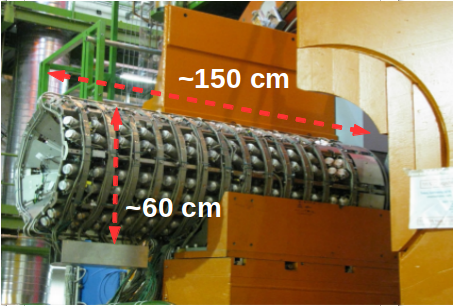}
\caption{}  
\label{detectors}
\end{center}
\end{figure}

{\bf Fig.~\ref{detectors}}. Left: Sketch of the CMS detector with specifications of the subdetectors; note the circle on the right to point the CASTOR calorimeter out. Right: Picture of the CASTOR calorimeter before the installation.

\begin{figure}[htbp]
\begin{center}
\includegraphics[width=8cm]{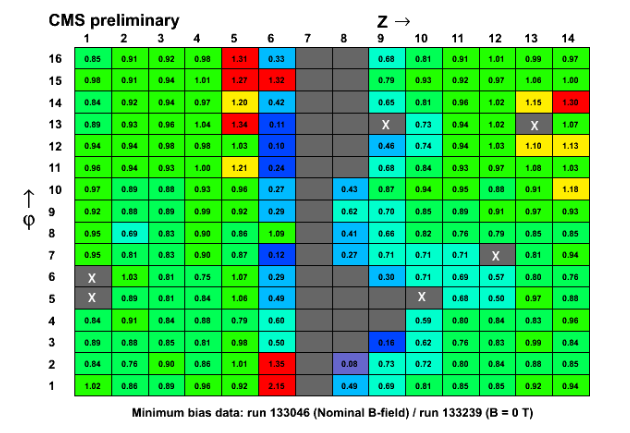}
\includegraphics[width=8cm]{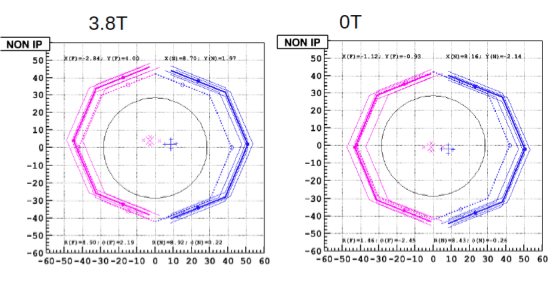}
\caption{}  
\label{magnfieldshifts}
\end{center}
\end{figure}

{\bf Fig.~\ref{magnfieldshifts}}. Left: Map ($\phi$ vs z) of the ratio $S_{i}$(B=3.8 T)/$S_{i}$(B=0 T) of the average response of all channels $i$ of CASTOR with and without magnetic field. The grey colour in the central region indicates a ratio close to 0, meaning a high inefficiency in presence of magnetic field, while the crossed channels have been observed to be dead regardless of the magnetic field. Right: Position of the rear side of the CASTOR half detectors at B=3.8 T and zero field.

\begin{figure}[htbp]
\begin{center}
\includegraphics[width=8cm]{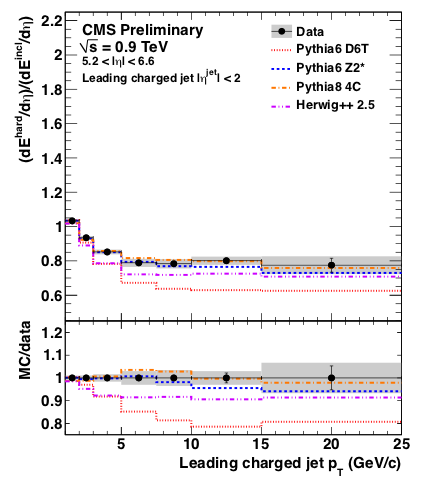}
\includegraphics[width=8cm]{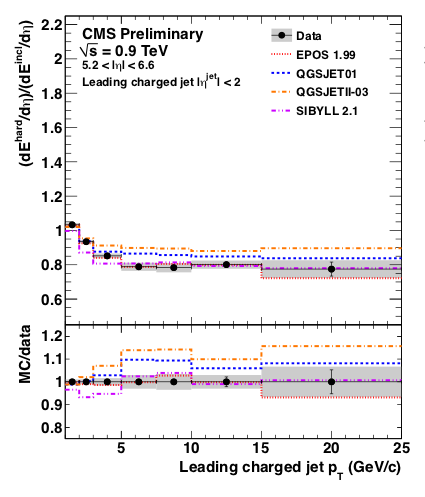}
\caption{}
\label{PPmeasurement1}
\end{center}
\end{figure}

{\bf Fig.~\ref{PPmeasurement1}}. Ratio of the energy deposited in the pseudorapidity range 5.2 $<$ $\eta$ $<$ 6.6 for events with a charged particle jet with $|\eta|$ $<$ 2 with respect to the energy in inclusive events, as a function of the charged particle jet $p_T$ for $\sqrt{s}$ = 0.9 TeV. The error bars indicate the statistical error while the band around the data points represent the systematic and statistical uncertainties added in quadrature.

\begin{figure}[htbp]
\begin{center}
\includegraphics[width=8cm]{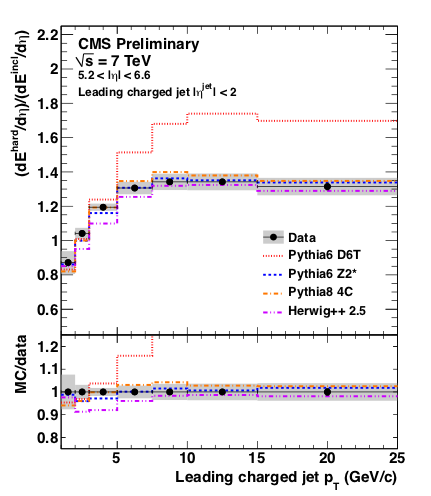}
\includegraphics[width=8cm]{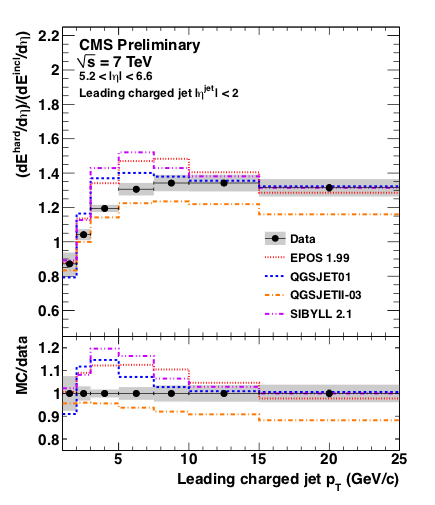}
\caption{}
\label{PPmeasurement2}
\end{center}
\end{figure}

{\bf Fig.~\ref{PPmeasurement2}}. Ratio of the energy deposited in the pseudorapidity range 5.2 $<$ $\eta$ $<$ 6.6 for events with a charged particle jet with $|\eta|$ $<$ 2 with respect to the energy in inclusive events, as a function of the charged particle jet $p_T$ for $\sqrt{s}$ = 7 TeV. The error bars indicate the statistical error while the band around the data points represent the systematic and statistical uncertainties added in quadrature.

\begin{figure}[htbp]
\begin{center}
\includegraphics[width=8cm]{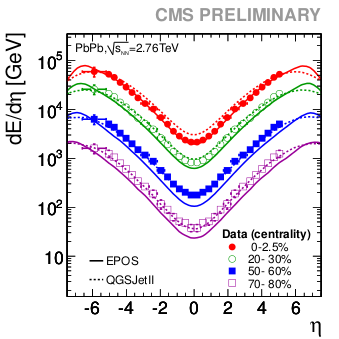}
\includegraphics[width=8cm]{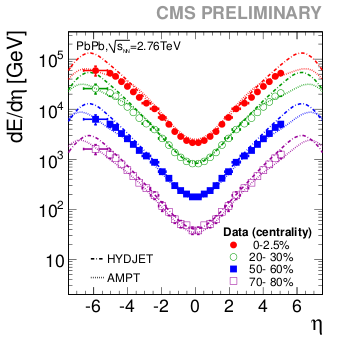}
\caption{}
\label{leadlead}
\end{center}
\end{figure}

{\bf Fig.~\ref{leadlead}}. Corrected energy density for different centralities. The vertical error bars are of systematic nature, while statistical uncertainties are too small to contribute.

\newpage

\end{document}